\documentclass[10pt, twocolumn]{IEEEtran}

\pdfoutput=1

\usepackage{hyperref}
\usepackage{color}
\usepackage{amsmath}
\usepackage{algorithm}
\usepackage{algorithmic}
\usepackage{array}
\usepackage{mdwmath}
\usepackage{eqparbox}
\usepackage{float}
\usepackage[font=footnotesize]{subfig}
\usepackage{fixltx2e}
\usepackage{stfloats}
\usepackage{amsfonts}
\usepackage{cite}
\usepackage{amsmath}
\usepackage{bbm}
\usepackage{amssymb}
\usepackage{mathrsfs}
\usepackage{amsbsy}	
\usepackage{graphicx}

\floatstyle{ruled}
\newfloat{algorithm}{tbp}{loa}
\floatname{algorithm}{Algorithm}

\long\def\hide#1{}

\newtheorem{theorem}{\bf Theorem}[section]

\newcommand{\droped}[1]{{\color{blue} \sout{}}}

\def\ba{\begin{array}}
\def\ea{\end{array}}
\newcommand{\beq}{\begin{equation}}
\newcommand{\eeq}{\end{equation}}
\newcommand{\bq}{\begin{eqnarray}}
\newcommand{\eq}{\end{eqnarray}}
\newcommand{\bqn}{\begin{eqnarray*}}
\newcommand{\eqn}{\end{eqnarray*}}
\newcommand{\bee}{\begin{enumerate}}
\newcommand{\eee}{\end{enumerate}}
\newcommand{\bi}{\begin{itemize}}
\newcommand{\ei}{\end{itemize}}

\begin{document}

\title{Impact of Energy Consumption on Multipath TCP Enabled Mobiles}

\author{\IEEEauthorblockN{Lily Minear\IEEEauthorrefmark{1} and Eric Zhang\IEEEauthorrefmark{1}}\\
 \IEEEauthorblockA{\IEEEauthorrefmark{1}Dept. of Information Technology and Electrical Engineering, ETH, Zurich.\\
Email: lily.minear@eeh.ee.ethz.ch, eric.zhang@eeh.ee.ethz.ch}}

\maketitle

\begin{abstract}
Multiple accesses are common for most mobile devices today. This technological advance opens up a new design space for improving the communication performance of mobile devices. Multipath TCP is a TCP extension that enables
using multiple network paths between two end systems for a single TCP connection, increasing performance and reliability. Meanwhile, when operating multiple active interfaces, multipath-TCP also consumes substantial more power and drains out bettery faster than using one interface. Thus, enabling Multipath TCP on mobile devices brings in new challenges.
In this paper, we theoretically analyze the underlying design choices given by the Multipath TCP. In particular, we theoretically formulate the relation between performance (throughput) and energy consumption for Multipath TCP. We find that sometime the throughput and energy consumption can be concurrently improved. 

\end{abstract}

\section{Introduction}

4G/LTE and WiFi accesses are common for most mobile devices today. This When multiple interfaces provide
connectivity to the Internet, a device is said to be ?multi-homed?,
i.e., it has several IP addresses and connects to the Internet topology
in different topological locations at the same time. When initiating
a communication session to another Internet host, a multi-homed
device can consequently choose between several paths to that destination
IP address. Even though they are equipped with multiple access interfaces, they only use one interface.  One access interface however may not be able to support the bandwidth requirement of emerging applications such as high definition video. It is therefore advantageous to enable concurrent use of the multiple interfaces.

Multipath TCP (MP-TCP) is a TCP extension that allows a TCP connection to stripe traffic over multiple interfaces. It is being standardized by the IETF \cite{IETF} and they are in active calling for proposal. Various aspects on MP-TCP have been investigated in the literature, e.g. \cite{FPKelly_MTCP,RSrikant,Damon,Peng2014MultipathTCP,ford2013multipath}. 
They mainly focus on general framework for Multipath TCP and not limited to mobile devices. 

Some of the earliest works including \cite{FPKelly_MTCP,RSrikant,Damon} focus on developing convergence results for fluid model. They show that Multipath TCP generally does not converge but the sum of the throughput for each individual will converge. This reveals the drawbacks of deploying Multipath TCP commercially. In \cite{Damon}, an inspiring algorithm is proposed that is verified in simulation that they can converge faster, even comparable with regular TCP. In \cite{Peng2014MultipathTCP}, a general framework is proposed that systematically analyze various performance tradeoffs of Multipath TCP.

People recently begin to look at how to deploy Multipath TCP on Mobile devices, e.g.\cite{wirelessall,wireless1,EWTCP,2011mobility,lim2014green,peng2014energy,chen2013measurement}. Many benefits of MP-TCP for mobile devices have been reported in the literature. In \cite{Damon}, they show that the throughput can be improved by using WiFi and 3G network together. In \cite{2011mobility}, they show that MP-TCP enables smooth handovers between 3G and WiFi. Thus, MP-TCP is promising in providing reliable and efficient connections for mobile devices by using connection diversity and resource pooling under dynamic environments. Please refer to more discussion on this pointer in \cite{perrucci2011survey,pqymptcp,charlie2012energy}

Deploying MPTCP to mobile devices introduces a new problem,
in other words, the additional energy consumption for operating
multiple network interfaces. Cell phones are frequently
constrained by the battery life, which is a crucial constraint on mobile devices. Processing power available continues
to grow exponentially, but battery storage increases slowly
by comparison. Thus, power consumption is an important
area of research, particularly in mobile devices such as smartphones.
In order to utilize MPTCP on mobile devices with
limited energy resources, it is important to understand the
power consumption behavior of MPTCP to ensure that it is
practical.

Various energy saving mechanisms have been proposed to reduce energy consumption when MP-TCP is implemented in mobile devices. In \cite{pluntke2011saving}, a scheduler based on Markov decision process is proposed. In \cite{2011mobility}, a mechanism that selects the most energy efficient path based on periodic probing of all paths are proposed. Moreover, since energy cost is the criterion for path selection, the throughput performance would be degraded significantly if the energy efficient path turns out to be the most congested path.

Our goal is to develop MP-TCP algorithms that can achieve both energy efficiency and good throughput for mobile devices. In particular, we are interested in developing a general framework that can generalize all the existing Multipath TCP algorithms. In \cite{peng2014energy}, they classify applications into two categories as follows:

 {\it Real-time applications}: The duration of a data transfer is usually fixed for this type of applications such as video streaming. The energy consumption is instantaneous power times duration of transfer. Since duration of transfer is usually determined by external factors and can be regarded as a constant, the energy consumption is proportional to the instantaneous power.

{\it File transfer applications}: The size of the data transfer is usually fixed for this type of applications such as file download. The energy consumption is instantaneous power times duration of transfer divided by transfer time.

Due to differences in properties of the real-time and file transfer applications, we propose a different algorithm for each application. We will see that  4G/LTE network provides the most energy efficient path for file transfer application while WiFi network provides the most energy efficient path for real-time applications. 

In this work, we propose a general framework that can be used to study energy consumption and throughput for MP-TCP enabled mobiles. We first examine 
MPTCP energy consumption
behavior via a combination of measurement, modeling, and
experimentation. We use measurement of power consumption from \cite{lim2014green} and regress them to determine conditions
under which MPTCP is more energy-efficient than either
standard TCP or MPTCP.

The remainder of this paper is organized as follows: Section
II provides the model and notations and
III presents utility maximization model proposed in \cite{FPKelly_TCP}. Section III proves theorems and our major results and we conclude this paper in section IV.

\section{Model and Notation}

\begin{table}
\caption{Key Notations}
\begin{center}
\begin{tabular}{|l|p{6.5cm}|}
\hline
Notation & Definition\\
\hline
$\mathbb{R}^n$           & n dimension real number\\
$\mathbb{C}^n$          & n dimension complex number\\
$\|x\|$        &  Euclidean distance of $x$\\
$N$    & Set of Users in the network\\
$R$   & Set of routers in the network\\
\hline
\end{tabular}
\end{center}
\label{tab:notation}
\end{table}

In this section, we introduce our notations and model for utility maximization proposed in \cite{FPKelly_TCP}, which will be used extensively in our paper. 

\subsection{Notations}

Let $\mathbb{R}^n$ denote the set of $n$ dimensional real number and $\mathbb{C}^n$ denote the set of $n$ dimensional complex number. $x\in\mathbb{R}^n$ means $x$ is a $n$ dimensional vector. Let
\begin{equation*}
\|x\|=\sqrt{\sum_{i=1}^{n}x_i^2}
\end{equation*}
represents the Euclidean length of a vector $x\in \mathbb{R}^n$, then $\|x+y\|\leq\|x\|+\|y\|$.

A function 
\begin{equation*}
U(x) \in \mathbb{R}\rightarrow \mathbb{R}
\end{equation*}
is defined as a {\bf Utility function} if $U(x)$ is a concave function of $x$.

A utility function that incorporates energy consumption is defined as 
\begin{equation}
U(x,e) \in \mathbb{R}^2\rightarrow \mathbb{R}
\end{equation}
defines users satisfaction of throughput at $x$ and energy consumption at $e$. 

\subsection{Utility function}

A function 
\begin{equation*}
U(x) \in \mathbb{R}\rightarrow \mathbb{R}
\end{equation*}
is defined as a {\bf Utility function} if $U(x)$ is a concave function of $x$.

A utility function that incorporates energy consumption is defined as 
\begin{equation}
U(x,e) \in \mathbb{R}^2\rightarrow \mathbb{R}
\end{equation}
defines users satisfaction of throughput at $x$ and energy consumption at $e$. 

$U(x,e)$ can take various form. The exact form of $U(x,e)$ depend on applications. Here
we list some of the common form of $U(x,e)$
\begin{itemize}
\item Linear form: this assumes users favor of a linear combination of throughput and energy consumption, e.g.
\begin{equation*}
U(x,e)=\alpha x+\beta e
\end{equation*}
\item Logarithmic  form: this assumes users favor of a linear combination of logarithmic of throughput and energy consumption, e.g.
\begin{equation*}
U(x,e)=\alpha \log(x)+\beta \log(e)
\end{equation*}
\item Negative quadratic form: this assumes users favor of a linear combination of negative quadratic form of throughput and energy consumption, e.g.
\begin{equation*}
U(x,e)=-\left(\alpha x^2+\beta e^2\right)
\end{equation*}
\end{itemize}

\subsection{Energy consumption model for mobiles}

\begin{figure}
\begin{center}
\includegraphics[width=0.4\textwidth]{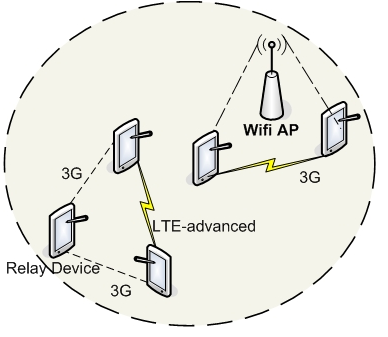}
\caption{Diagram of Multipath TCP for Mobile devices. The cellphones(mobiles) has access to both LTE network and WiFi network.}
\label{default}
\end{center}
\end{figure}

The 3rd Generation Partnership Project standard
redefines a state machine for the 4G/LTE communication interface,
which describes the possible power states of each device
connected to the network. To initiate a packet transmission,
a 4G/LTE communication interface(access point)\footnote{An access point consists of a base station and an end device.} has to switch from a very low
power to high power state so that packet can be transmitted
or received. If there is no further packet transmission for a
certain amount of time, to save energy, the 4G/LTE access will return
to a low power.

\section{Analysis}

In this section, we will show that it is theoretically possible to enhance throughput and energy consumption at the same time. However, it is nontrivial to design an algorithm that archives such good performance and we will leave it as our future work. This is similar to Shannon's information theory that developing a realizable codes that actually archives the information upper bound is NP hard.

\subsection{Formulation}
In this subsection, we will develop an optimization problem that can help us understand the design of mutlipath TCP algorithms for mobile devices. 
The objective is usually a sum of utility function of users' satisfaction of the throughput and energy consumption, i.e., we will minimize
\begin{equation*}
\sum_{i\in N }U_i(x_i,e_i),
\end{equation*}
where $x_i$ and $e_i$ represents user $i$'s throughput and energy consumption. 

We consider the following constraints:
\begin{itemize}
\item The relation between rate $x_i$ and $e_i$. The energy consumption is usually a nonlinear function of the throughput, i.e. 
\bqn
e_i=f_i(x_i) \quad \text{ for some particular form of $f_i$}
\eqn
\item The battery constraints for a mobiles. The total energy consumption for a user is upper bounded by its battery life $B_i$, i.e.
\bqn
\sum_{t=0}^Te_i\leq B_i \quad i \in N
\eqn
\item The capacity constraints for a mobiles. The total capacity within a cell $c$ should be upper bounded by $X_c$,i.e.
\bqn
\sum_{i\in N\cap c}\mathbb{P}(x_i,c)\leq X_c \quad c\in C
\eqn
\item The protection of users' battery life. The battery cannot work at high power level for two long time. We use a decaying causal filter to model this aspect, i.e. 
\bqn
\sum_{t=0}^Te_i(t)d^t\leq D
\eqn
\end{itemize}

Then the optimization problem we will solve takes the following form:
\begin{eqnarray}
\min &&  \sum_{i\in N }U_i(x_i,e_i) \\
s.t.  &&  e_i=f_i(x_i) \\
&& \sum_{t=0}^Te_i\leq B_i \quad i \in N\\
&& \sum_{i\in N\cap c}\mathbb{P}(x_i,c)\leq X_c \quad c\in C\\
&& \sum_{t=0}^Te_i(t)d^t\leq D\\
&& x_i\leq \overline{x}_i\\
&& e_i\leq \overline {e}_i
\end{eqnarray}

\subsection{Analysis}
The solution to our optimization problem determines the optimal structure of allocating energy consumption for each path for  a mobile device. 
\begin{theorem}\label{theorem1}
The optimization problem (3) is a NP-hard problem if the realization of $f_i(x_i)$ is nonlinear. 
\end{theorem}
\begin{IEEEproof}
A decision problem D is NP-hard when for any problem A in NP, there is a polynomial-time reduction from D to A. An equivalent definition is to require that any problem D in NP can be solved in polynomial time by an oracle machine with D. Generally speaking, we can regard an algorithm that can call such an oracle machine as a subroutine for solving D, and solves A in polynomial time, if the subroutine call takes only one step to compute.

If $f_i$ is linear, then the problem (2) is a LP problem, which can be solved by either simplex method or interior point method. When $f_i$ is nonlinear, the problem will be NP-hard to solve. People can refer to \cite{boyd2004convex} for more details on nonconvex NP hard problem.  
\end{IEEEproof}
Theorem \ref{theorem1} says that solving problem (2) is nontrivial even in a centralized manner. Thus, it is hard to give an efficient algorithm that can actually solves the MP-TCP enabled mobile devices problem.


\section{Conclusion}
In this paper, we theoretically formulate an optimization problem that can help determine the optimal energy allocation for Multipath-TCP enabled mobile devices. We theoretically prove that the problem is NP hard due to the nonlinear property of the relation between energy consumption and throughput. We 
conjecture that it is possible to improve both performance and throughput concurrently by using multi path TCP on mobiles.

\bibliographystyle{IEEEtran}
\bibliography{IEEEabrv,reference}


\end{document}